\documentclass[%
reprint,
amsmath,amssymb,
aps,
]{revtex4-2}

\usepackage{graphicx}
\usepackage{dcolumn}
\usepackage{bm}

\usepackage[utf8]{inputenc}
\usepackage[T1]{fontenc}
\usepackage{mathptmx}
\usepackage{etoolbox}
\usepackage[version=4]{mhchem}
\usepackage{xcolor}
\usepackage{physics}

\makeatletter
\def\@email#1#2{%
\endgroup
\patchcmd{\titleblock@produce}
{\frontmatter@RRAPformat}
{\frontmatter@RRAPformat{\produce@RRAP{*#1\href{mailto:#2}{#2}}}\frontmatter@RRAPformat}
{}{}%
}%
\makeatother

\begin{document}

\preprint{APS/123-QED}

\title{Simulating Mono- and Multi-Protein Phosphorylation within Nanoclusters}
\author{Olivier Destaing}
\affiliation{
Institute for Advanced Biosciences, Centre de Recherche Universit\'e Grenoble Alpes, \\ Inserm U1209, CNRS UMR 5309, France.
}
\author{Bertrand Fourcade}
\email{Bertrand.Fourcade@univ-grenoble-alpes.fr}
\affiliation{%
Universit\'e Grenoble Alpes and CNRS, LIPHY, F-38000 Grenoble, France}
\date{\today}

\begin{abstract}
Protein nanoclustering is a characteristic feature of their activated state and is essential for forming numerous subcellular structures. The formation of these nanoclusters is highly dependent on a series of post-translational modifications, such as mono- and multi-phosphorylation and dephosphorylation of residues. We theoretically simulate how a protein can be either mono- or multi-phosphorylated on several residues in functional nanoclusters, depending on effective biophysical parameters (diffusion, dwell time, etc.). Moving beyond a binary view of phosphorylation, this approach highlights the interplay between mono- and multi-phosphorylation, the cooperative effects generally associated with multi-phosphorylation networks, and stresses the role of phosphatases in transforming graded phosphorylation signals into almost switch-like responses. The results are discussed in light of experiments that probe the distribution of phospho-residues.
\end{abstract}

\maketitle
\section{\label{sec:introduction} Introduction}
Living cells process information from their extracellular environment through biochemical signaling networks, where protein phosphorylation is one of the mechanisms responsible for altering their biological activity, location, or docking with other proteins \cite{Marks:2009yo}. These regulatory mechanisms are orchestrated by a large set of kinases and phosphatases, where a protein can be mono- or multi-phosphorylated on different residues \cite{Cohen:2000ty,Koivomagi:2013bl}. Protein phosphorylation is a local and contextual phenomenon, meaning the same system of molecular components can display different properties in different situations. 

An emerging concept in many areas of biology, including cell and developmental biology or immunology, is the existence of finite-lifetime platforms at sub-resolution length scales where proteins are phosphorylated \cite{Gormal:2024xy, Remorino:2017os, Zhou:2015oh, Garcia-Parajo:2014ua, Rossy:2012iq, Kholodenko:2010ty,Bray:1998kq}. This non-homogeneous spatial organization aligns with reports showing 40-250 nm nanoclusters of different adhesome proteins throughout adhesions, which organize these protein covalent modifications for signal and information transmission \cite{Changede:2015rp,Bachmann:2022tu}.

Following classical works that established the first-passage framework \cite{Berg:1977ul,Shoup:1982si,redner}, this paper reformulates these concepts within reaction-diffusion models for nanoclusters to show how essential properties can be combined \cite{Kholodenko:2010ty, Collinet:2021rw}. The first, called   pleiotropy, refers to the possibility for the same signaling element—a kinase, phosphatase, or GTP-ase, for example—to induce different responses. The second is the possibility of  crosstalk   between signaling pathways, where one signaling input is amplified or switched off depending on a second one. The third is the specific   decision-making   that explains how a single signaling event can generate different cellular responses. Although discussed in the biological literature since the 1980s and in physics more recently \cite{Kirby:2021ix}, the decision-making process between antagonistic pathways has mainly been studied at the molecular level in a binary representation where a protein undergoes post-translational modifications \cite{Hynes:2013qg}. While important, these works provide only a limited understanding of experiments where, for example, optogenetic control of signal processing modifies and selects cellular responses \cite{Toettcher:2013qy,Farahani:2021lu,Kerjouan:2021ml,Ju:2022ub,Seze:2025ik}. Model systems that combine molecular data with biophysical constraints could therefore be useful in a context where changing the dynamics of a single molecule elicits drastic changes, as demonstrated by  spatio-temporaly resolved phospho-proteomics\cite{Toettcher:2013zi, Zhou:2023ew}. 

As applied here, bidimensional ligand-receptor networks at the front-end of cell signaling system are complex signal-units able to select among different pathways  for different environment\cite{Vilar:2006vz,Antebi2017}. Since phosphorylation by kinases is a key regulatory layer for protein-protein interaction networks, these experiments and these conceptual approaches  prompt the inclusion of 2D diffusion effects. By including diffusion into effective nanoscopic phosphorylation rates, the aim here is to show, from the perspective of mono-and-multiphosphorylation, how the following questions can be answered   
 : How is a substrate selected? What is the probability for a protein to be phosphorylated? What is the phosphorylation mode, and how does its likelihood change with varying parameters? Is it an all-or-nothing process, an almost switch-like process, or a graded rheostat-type process? Answering these questions is a first step toward a better understanding of a system capable of different molecular self-organization using the same set of proteins \cite{Saiz:2006mj,Albiges-Rizo:2009bq}. We will answer this question here by calculating exactly  the 2D diffusive flux of a molecule with different boundary conditions, as this problem has already been addressed in Ref. \cite{SHOUP1982237}.

Biophysical models for mono- and multi-phosphorylation in signaling to model information processing have been the subject of previous works based on different versions of mass-action kinetics \cite{Swain:2002nq,salazar2009av, Fran_ois_2016, alon}. They are generally based on the McKeithan model \cite{McKeithan:1995ph}, which generalizes the Hopfield-Ninio paradigm \cite{Hopfield:1974lr, Ninio:1975fk}. In these models, an enzyme-bound substrate must complete a series of phosphorylation events for a cellular response to occur, see \cite{kirby2025} and references therein. Here, the spirit is the same, but the molecular kinetic constants are replaced by effective parameters that allow the system to probe biophysical constraints, such as the size of clusters, the dwell time of proteins specifically anchored on the plasma membrane, or their diffusion, which can vary from one zone to another. Our approach differs from that in other references, such as \cite{Gopich:2016xe,Gopich:2024rw}, which assume an escape probability for diffusion to modify the kinetic rate constants in 3D. Since random walks are recurrent in 2D \cite{redner}, we take a different approach and we solve the full diffusion time-dependent diffusion problem from an analytical and stochastic point of view. Crosstalks between different pathways can then be naturally integrated into this approach where one signaling pathway reads local, averaged biophysical properties modified by another. This spatiotemporal approach also takes into account diffusive and rebinding effects, which are known to induce significant changes in molecular complexes produced in response to the phosphorylation of substrates, leading to changes in the signaling network \cite{Takahashi:2010nr, Mugler:2012oq, Gopich:2013sc,Mugler:2012oq,Mugler:2013qy} or to the induction of biomolecular condensates \cite{Huang:2016oz, Huang:2019rm, Case:2019oy, Oh:2012ri}.

This paper focuses on the calculation of the probability for a protein to be mono- or multi-phosphorylated. Multi-phosphorylation is crucial since it can involve a large number of states whose stability may depend on this number \cite{Bialek2000,Warren:2005vj}. Thanks to its combinatorial properties, the present paper shows that the process of multi-phosphorylation combined with its reverse process leads to very different responses depending on the averaged local environment, reflecting the presence of different signaling pathways. Multi-phosphorylation also allows for cooperativity. In cell signaling, cooperativity is often defined as the result of an allosteric phenomenon, where a chemical change in one of a protein's residues (in this case, phosphorylation) affects the affinity of another residue for another reaction. However, multiple binding can also be induced by multi-phosphorylation-sustained cooperativity, a process often referred to as   avidity   \cite{Erlendsson:2020kl}, which we know can lead to the phenomenon of   superselectivity   due to entropic factors \cite{Albertazzi:2013lw,Curk:2017gf,Dubacheva:2019df}. Because of this avidity effect, bivalent induced phospho-binding with proper spacing, even contributed by two weak bonds, can produce stable protein complexes and may be important for understanding the specificity of some signaling phosphatases \cite{Lim:2002ts,Xu:2021zy}. We show here that phosphatases, when integrated into the model, have strong effects on the calculation of these probabilities, with potential switch-like effects.

In short, for a relatively small number of residues, the model presented below demonstrates that multi-phosphorylation leads to new phenomena including, for example, cooperativity, changes in critical time scales, and graded versus almost switch-like phosphorylation modes. Steep variations and abrupt changes in phosphorylation likelihoods have the potential to have consequences on the biological side, and this work highlights the importance of having multiple interaction sites for phosphorylated substrates, as described by the modular logic for some kinases and phosphatases \cite{Lim:2002ts}.

We proceed both analytically, within a type of effective medium approximation, and numerically. Numerical simulations using a spatial Gillespie's algorithm \cite{Gillepsie2007,Erban:2009aa} are useful because they provide access to the stochastic regime for systems with very few molecules and to quantities that are difficult to obtain analytically. The method is standard and is briefly summarized in Appendix \ref{Appendix0}. This paper is structured as follows. The first sections introduce the notations and specify the various models. The case of mono-phosphorylation will then be discussed. Multi-phosphorylation and, finally, the effect of phosphatases will be discussed last. The conclusion summarizes these elements and briefly addresses the potential biological applications, while technical details can be found in the appendices.

\section{The Models}

In parallel with Single Particle Tracking experiments that probe the journey of a fluorescent protein within integrin-based structures (see \cite{Rossier:2016zr,Orre:2021we}, for example), we model a process where a protein is deposited at time $t=0$ anywhere within an activation disk of size $R_a$. Efficient binding to a partner requires mono- or multi-phosphorylation of key residues. The probability of success of this process and its average time are assumed to be measurable. To simulate this process, we attach a virtual counter to this protein and stop the process when a given number $n$ of residues is phosphorylated, with $n=1$ for mono-phosphorylation and $n>1$ for multi-phosphorylation. Having a given number of phospho-residues is a specific outcome out of several alternative possibilities, since the protein can desorb from the membrane and return to the cytosol or be absorbed on specific boundaries. This is an example of splitting probability where the process  can split into two different outcomes\cite{redner}.

The physical framework presented here treats free diffusion along the membrane and efficient phosphorylation as two generic processes. In addition to phosphorylation, diffusion along the membrane is also a regulatory layer, since in-plane diffusion requires proteins to adhere to the membrane through specialized protein domains. The model, therefore, considers diffusion coefficients, which may vary depending on the membrane domain, and the dwell time of proteins on the membrane as parameters, since not all proteins are anchored in the same way, as shown in \cite{Orre:2021we}, for example.

\begin{figure}[!htb]
\includegraphics[scale =0.3]{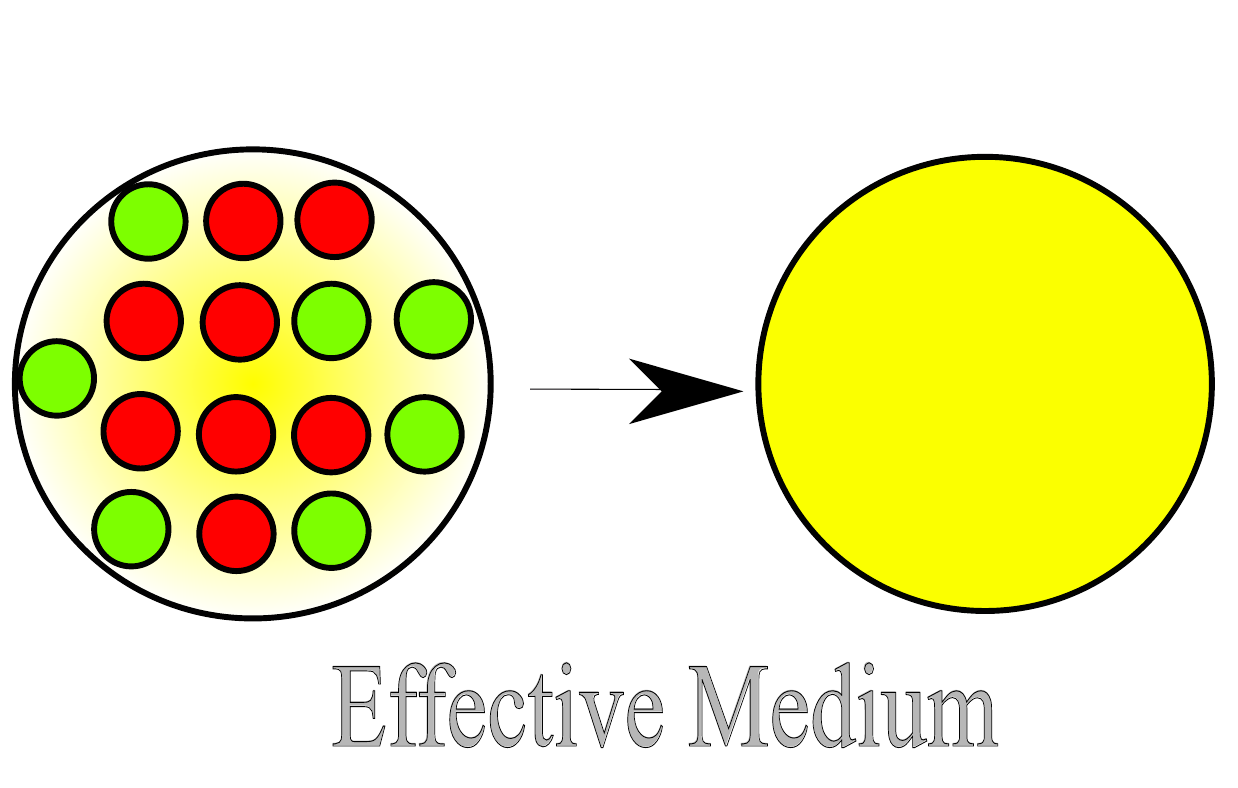}
\caption{\label{fig:fig1}An activation disk is a region divided into elementary cells where a protein can diffuse with diffusion coefficient $D = D_1$ and be phosphorylated at a rate of $\mu_p$. Each cell can oscillate over time between two states: active ($\mu_p >0$), red (light gray) symbols, and inactive ($\mu_p=0$), green (dark gray) symbols. This description is suitable for stochastic calculations. For the analytical approach, the effective medium approach means averaging the phosphorylation rate over the entire activation disk (yellow disk). Dephosphorylation will be introduced in a similar way with a local dephosphorylation rate $\mu_{dp}$.
}
\end{figure}

At the nanoscale, as seen in Fig. \ref{fig:fig1}, adhesive structures are segregated into distinct functional nanoclusters or nano-aggregates, some of which are elementary units where a protein can be phosphorylated on specific residues. From now on, we refer to a nanocluster as the elementary functional unit where a protein can be phosphorylated at some rate $\mu_p$. In a numerical analysis, where the two-dimensional space is partitioned into units, this unit is referred to as an elementary cell. Each elementary cell models a protein kinase nanocluster, and the set of elementary cells is the activation disk, loosely representing adhesive structures. The size of the activation disk, or the number of elementary units, is a parameter of the model, and their number ranges from a few to dozens. In short, the disk is a functional model for a signaling platform with a group of immobilized kinases.

\begin{figure}[!htb]
\includegraphics[scale =0.7]{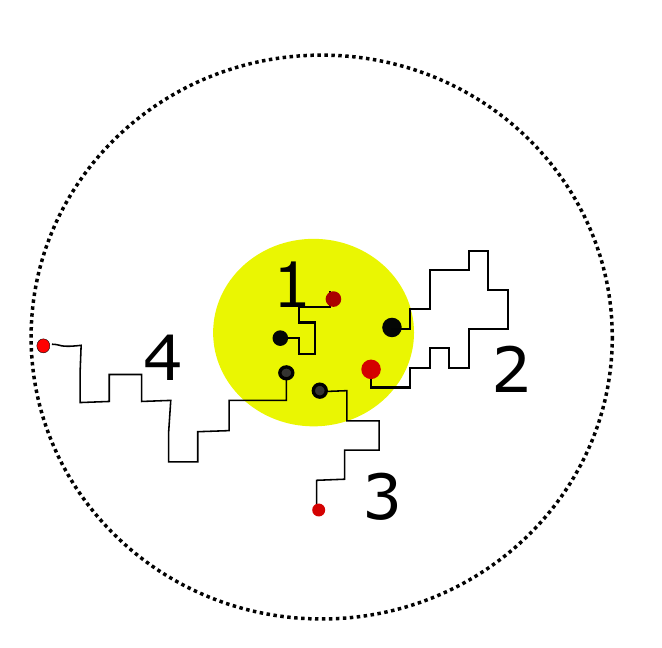}
\caption{\label{fig:fig2}Four elementary processes for a protein initially deposited in the activation disk (yellow, see Fig. \ref{fig:fig1}) with diffusion coefficient $D_1$ inside the disk and $D_2$ outside: (1)
The protein is phosphorylated before exiting the disk; (2) The protein is phosphorylated after a round trip (loop effect);
(3) The protein is desorbed from the membrane outside the disk. Event (4) with the outside dashed circle only concerns numerical simulations and can be set aside on first reading, see Appendix \ref{Appendix0}. }
\end{figure}

Proteins enter and exit by 2D diffusion, and  they can be (de)phosphorylated at some rate ($\mu_p$ and $\mu_{dp}$), as shown in Fig. \ref{fig:fig2}. These rates are parameters of the model and are proportional to the catalytic rates in the non-saturated regime \cite{alon}. They can also vary over time due to a fast-varying control variable, such as a change in conformation. This is the dynamic disorder case, a problem tackled here from a numerical perspective. For analytical calculations, we will use the effective medium approximation, which here consists of averaging the phosphorylation rate over all cells. This approximation holds if $\mu_p$ fluctuates sufficiently rapidly, and we show how this process limits the accuracy of this approximation in the probability calculation.

After the start signal at $t=0$, the process ends in two cases: either the protein is phosphorylated, or it desorbs from the membrane at a rate $\mu_d$. What is the probability of phosphorylation? What is the average phosphorylation time? Note that if we allow loop trips by diffusion between the disk and its outside, the random walk being recurrent in 2D \cite{redner}, we must impose a finite rate of desorption from the membrane to the cytosol; otherwise, this probability is 1. For a finite desorption rate, the protein is said to be mortal \cite{Meerson:2015pb}.

To better understand the role of the parameters, it is useful to first consider i) a constant diffusion constant $D$ along the membrane and ii) the two limiting cases for the completion time probability:
\begin{enumerate}
\item   For immortal proteins (model A):   We impose absorbing boundary conditions with zero probability at the disk boundary $r=R_a$. The process is complete and stops when the protein is phosphorylated (success) or when it reaches the boundary of the disk (failure). The problem is characterized by two length scales: the activation length, $l_p = \sqrt{D/\mu_p}$, and $R_a$. Since the solution to the problem involves dimensionless factors, the ratio $l_p/R_a$ is the only parameter. To within a numerical factor, this ratio is the square root of the diffusion time over the average phosphorylation time at a single site.
\item   For mortal proteins (model B):   The process comes to an end when either the protein is phosphorylated (success) or it desorbs from the membrane (failure). The desorption length $l_d=\sqrt{D/\mu_d}$ is the third length scale. We assume $l_d >R_a$. A mortal protein can perform many round trips before being phosphorylated if its desorption rate is sufficiently small. This is a loop or round-trip effect.
\end{enumerate}

By analogy with a biological context, the differences between the two models can be visualized as follows. The first corresponds to a protein recruited from the cytosol that can bind other elements in the signaling platform (as observed for proteins able to both bind other proteins and the 2D slowly diffusing membrane of the platform). The second corresponds to the more general situation where the protein can return many times to the activation disk before desorbing into the cytosol.

Until now, we have assumed equal, constant diffusion coefficients inside and outside the activation disk. This approximation is questionable since some proteins can interact with multiple partners before finding their target following phosphorylation. The problem is generally solved using an effective diffusion constant. This point is considered in a third section, where we show that using different diffusion coefficients $D_{1,2}$ for the inside and outside of the disk enables a crossover between the two limiting cases of models A and B above.


\section{The mono-phosphorylation case }
\begin{figure}[!htb]
\includegraphics[scale =0.7]{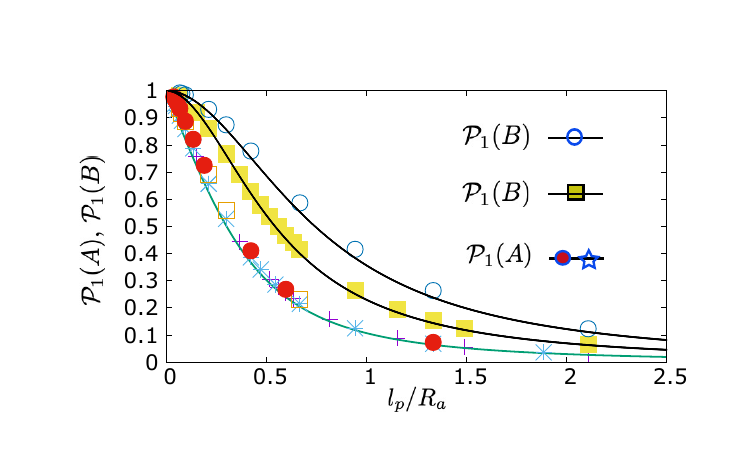}
\caption{\label{fig:fig3}Probability of a protein being phosphorylated once as a function of the parameter $l_p/R_a$, where $l_p/R_a \ll 1$ corresponds to the strong phosphorylation regime. The bottom curve corresponds to model A (immortal protein) with absorbing boundary conditions at the boundary $r=R_a$, see (\ref{eq:proba_one_model_A}). Points correspond to numerical data with different values of $D, \, \mu_p$ and $R_a$ to illustrate how different systems collapse on the same curve. The two upper curves correspond to mortal proteins (model B, see (\ref{eq:proba_one_model_B})) for two values of $R_a=3$ (upper curve, $l_d/R_a\simeq 1.89$) and $R_a=6$ (median curve, $l_d/R_a \simeq 0.94$). The smaller the radius $R_a$, the greater the effect of round trips (loop effect). }
\end{figure}

In both cases, immortal and mortal proteins, not all proteins are phosphorylated after random deposition in the disk. In the mono-phosphorylation case, the calculation amounts to determining the propagator of a diffusion equation
\begin{equation} \label{eq:first_equation}
\frac{\partial {\mathcal{P}}_1}{\partial t} = D \Delta {\mathcal{P}}_1 - \mu_{1,2} {\mathcal{P}}_1(r,t)
\end{equation}
with the initial condition
\begin{equation}
{\mathcal{P}}_1 (r,t=0) = \frac{1}{2 \pi r_0 }\delta(r-r_0)
\end{equation}
and where the subscript 1 refers to the inside of the disk, $\mu_1= \mu_d + \mu_p \simeq \mu_p$, and 2 to the outside, $\mu_2 = \mu_d$. To calculate the probability ${\mathcal{P}}_1$ to be phosphorylated once, one calculates
\begin{equation}
{\mathcal{P}}_1 = \frac{2 \mu_p}{R_a^2} \int_0^{R_a} dr_0\,r_0 \int_0^\infty dt {\mathcal{P}}_1(r_0,t)
\end{equation}
to average over the starting point $0 \le r_0 \le R_a$.

\subsubsection{Model A}
Using the calculations in Appendix \ref{Appendix1}, it is first interesting to discuss the case of an immortal protein with absorbing boundary conditions (model A). As pointed out before in other contexts \cite{krapivsky2017,Berezhkovskii2019}, this probability is a function of only the dimensionless ratio $l_p/R_a$
\begin{equation} \label{eq:proba_one_model_A}
{\mathcal{P}}_1(A) =1 - 2 \frac {l_p}{R_a} \frac{I_1(R_a/l_p)}{I_0 (R_a/l_p)}
\end{equation}
where $I_{0,1}$ are modified Bessel functions, as shown in Fig. \ref{fig:fig3}, and where the strong phosphorylation regime corresponds to small values of the parameter $l_p/R_a$.

This scaling can be made more explicit by noting that the ratio
$
l_p/R_a = \sqrt{D/(\mu_pR_a^2)}
$
is, within some numerical factor, the square root of the fraction of time it takes for the protein to be phosphorylated without outflow to the typical time it takes for a free protein to diffuse out of the activation disk. Let $N_a$ be the total number of kinase proteins in the disk. By definition, $\mu_p$ is an average rate and $\mu_p = N_a k_a h^2/R_a^2$, where $k_a$ is the production rate per enzyme of the product in the Michaelis-Menten kinetics \cite{alon} and $h$ the size of a functional enzyme. Thus $l_p/R_a$ depends on $N_a$ but not on $R_a$. As a consequence, the scaling plot in Fig. \ref{fig:fig3} shows that the probability for phosphorylation is independent of the size of the activating disk for a given number of kinases, because increasing the concentration by decreasing the size does not change the ratio of these two time scales.

However, because diffusion takes time, the size of the disk does matter for the mean phosphorylation time $<t_{phos}>$, and it increases with the size $R_a$. For future reference, the natural scale for this process is the mean escape time through diffusion from the disk, $t_{diff} = R_a^2/8D$, where the factor 8 arises because one averages over the starting point.

To calculate this time, we refer to the calculation in Appendix \ref{Appendix3}. The probability ${\mathcal{P}}_{1} (A,t) dt$  of the protein being phosphorylated once between $t$ and $t+ dt$  is 
\begin{equation} \label{eq:probab_model_A}
{\mathcal{P}}_{1} (A,t) dt = \frac{1}{{\mathcal{Z}}} \mu_p e^{-\mu_p t} (1- {\mathcal{F}}_{\mu_p=0}(t)) dt
\end{equation}
In (\ref{eq:probab_model_A}), ${\mathcal{F}}_{\mu_p=0}(t)$ is the probability to cross the boundary of the disk for the first time at exactly $t$, and ${\mathcal{Z}}$ is a normalization factor.

\begin{figure}[!htb]
\includegraphics[scale = 0.7]{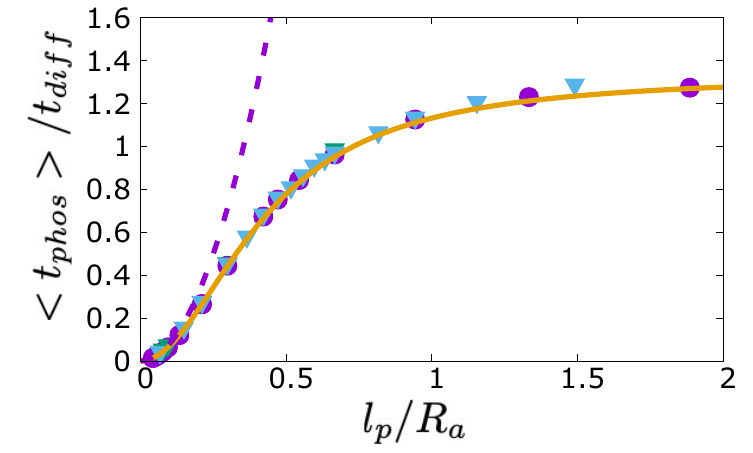}
\caption{\label{fig:mean_time} Average time for a protein to be phosphorylated at a single site, normalized by diffusion time, see (\ref{eq:first_moment_resu}). As indicated by the points obtained numerically, this curve, once normalized, is a scaling curve as a function of $l_p/R_a$. The dotted curve corresponds to $<t_{phos}> = 1/\mu_p$, which would be the value of the mean phosphorylation time for a disk of infinite radius. Down triangles and circles correspond to different sizes $R_a$ and make the scaling property explicit. For $l_p/R_a >1$, the probability that a protein will be phosphorylated is negligible.}
\end{figure}

Taking the average value over the subset of proteins that have been phosphorylated exactly once, one finds
\begin{equation}
<t_{phos}> = t_{diff} {\mathcal{T}}(l_p/R_a)
\end{equation}
where the function ${\mathcal{T}}(l_p/R_a)$ is given in Appendix \ref{Appendix3}, see (\ref{eq:first_moment_resu}), and is plotted in Fig. \ref{fig:mean_time}. This calculation shows that $<t_{phos}>$ is generally smaller than $1/\mu_p$, with equality only in the very high phosphorylation rate regime. This trend can be explained by the fact that diffusion limits phosphorylation over short periods of time and skews the sampling to calculate the average by imposing a cut-off for the long times.

\begin{figure}
\includegraphics[scale= 0.7]{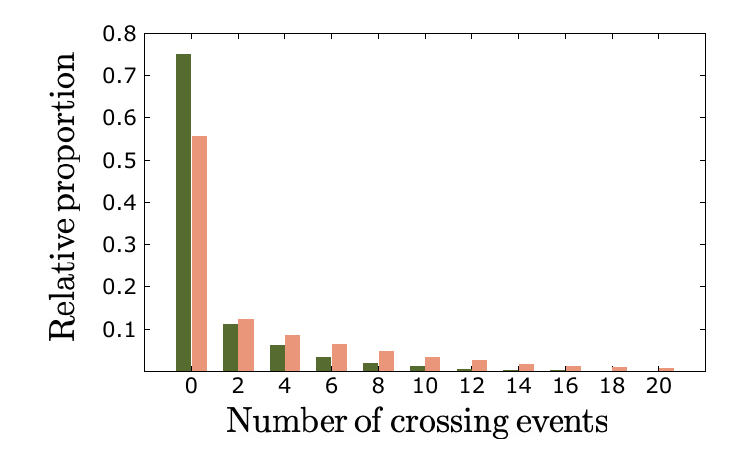}
\caption{\label{fig:fig5}
Normalized histograms of the number of times a protein crossed the boundary of the adhesive disk before being phosphorylated. There are therefore only even values. The two plots correspond to $l_p/R_a=0.21$ (orange - light gray) and $0.42$ (green - dark gray). For small $l_p/R_a$, in the high phosphorylation rate regime, proteins are rapidly phosphorylated. Data corresponds to $\mu_d = 5 \times 10^{-4}, \, R_a=3$ so that $l_d/R_a=1.89$. }
\end{figure}

\begin{figure}
\includegraphics[scale= 0.7]{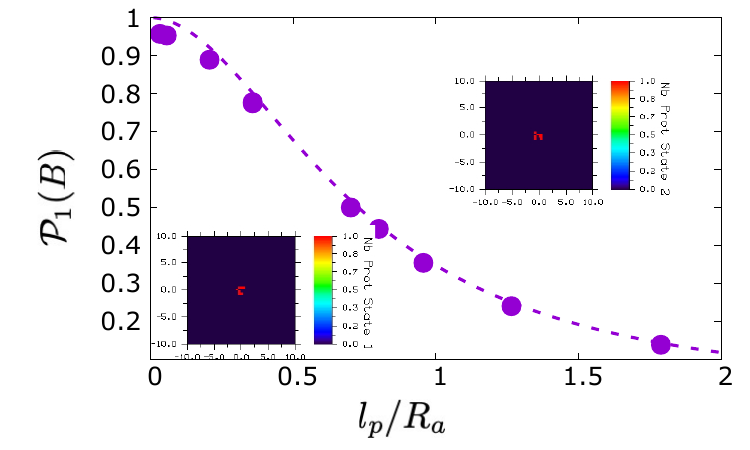}
\caption{\label{fig:fig6} Dynamic disorder case: probability of a protein being mono-phosphorylated (model B) when the activation zone in Fig. \ref{fig:fig1} contains only 16 cells dynamically alternating between two states, one of which can phosphorylate and the other not. For this small cluster, the activation zone has the shape of a square, as seen in the insets which represent snapshots of the two configurations in the activation zone. The stochastic simulation runs with equal rates $k_+ = k_-$ between the two states. The points with symbols $\bullet$ correspond to slow rates (0.01s$^{-1}$), while the dashed curve corresponds to (\ref{eq:proba_one_model_B}) with an effective rate $\mu_p/2$. }
\end{figure}

\subsubsection{Model B: Homogeneous case}

The specific case of model A is in stark contrast with the general result where we allow loop effects (model B) for mortal proteins ($D_1=D_2=D$). Referring again to Appendix \ref{Appendix1}, the general result reads as
\begin{equation} \label{eq:proba_one_model_B}
{\mathcal{P}}_1 (B) = 1 - 2 \frac{l_1}{l_d} \frac{I_1(R_a/l_1) K_1(R_a/l_d)}{ R_a Z_1}
\end{equation}
where $Z_1$ is a combination of Bessel functions and $l_1 = \sqrt{D/(\mu_d + \mu_p) }\simeq l_p, \, \mu_d \ll \mu_p$. As it should, result (\ref{eq:proba_one_model_A}) follows from (\ref{eq:proba_one_model_B}) in the limit of a high desorption rate outside the activating disk, $l_d \rightarrow 0$. As shown in Fig. \ref{fig:fig3}, this probability increases markedly, even for moderate values of desorption length, indicating the importance of round-trip effects.
This increase in probability naturally depends on the average time that must elapse before the first phosphorylation occurs. This time, in turn, depends on the desorption rate. In the case of model B, the reference time for this process is set by the average time $t_{dwell}$ that the protein has spent in the disk when diffusing back and forth.
Calculations in Appendix \ref{Appendix:time_spend} show that $t_{dwell}$ is given by

\begin{equation}
t_{dwell}/t_{diff} = \frac{l_d^2}{8R_a^2} \left[ 1 - 2 \frac{R_a}{l_d} I_1 (R_a/l_d) K_1(R_a/l_d) \right]
\end{equation}
with rapid variations when increasing $l_p/R_a$. A simple way to quantify this loop effect is to count the number of times a protein needs to cross the boundaries of the disk back and forth before being phosphorylated. The result is shown in Fig. \ref{fig:fig5} in the high and moderate phosphorylation regime, with the conclusion that even for moderate ratios $l_d/R_a$, the effect of allowing loops is strong due to the long tails common in these kinds of problems. The importance of these tails is particularly apparent in Fig. \ref{fig:fig5}, which shows how a protein can cross the disk boundary some ten times, even though the desorption length is only twice the size of the disk. This increase in probability is naturally at the expense of the average phosphorylation time, which is greater than in the previous case.

\subsubsection{Model B: Dynamic disorder case}
Fig. \ref{fig:fig5} is obtained using a stochastic solver for the Master equation equivalent to the problem \cite{fourcade:hal-02465425} and Appendix \ref{Appendix0}. The advantage of the numerical approach is the ability to compare the previous results with the dynamic disorder case, where the disk has a very limited number of kinases that alternate with rates $k_+, \, k_-$ between an active state, capable of phosphorylating the protein with $\mu_p>0$, and an inactive state, unable to do so with $\mu_p=0$. In other words, we assume that the enzymatic activity per cell depends on a time-varying control parameter ${\mathcal{C}}$, such as a change in molecular conformation, as in the classical gated binding problem \cite{McCammon:1981nx, zwanzig1990,Gopich:2016bq}. This telegraphic noise, often called dichotomous noise, breaks detailed balance and is often used to describe non-equilibrium states\cite{BENA_2006,smith2022}.

For fast time-varying control parameters, we expect renormalized effective rate constants. Fig. \ref{fig:fig6} plots the probability ${\mathcal{P}}_1 (B)$, where $l_p/R_a$ is now a function of the time-averaged enzymatic activity defined from the fraction of time spent in the active conformation:
\begin{equation}
\mu_p \rightarrow <\mu_p> = k_+/(k_+ + k_-)\mu_p
\end{equation}
If the fluctuations are sufficiently fast, it is sufficient to take the time average $<\mu_p>$, and the probability coincides with (\ref{eq:proba_one_model_B}). In the other limit of static disorder for slow variations between the active and inactive state, the two curves differ slightly, because for a low number of kinases, there is a non-zero probability that the protein will desorb without being phosphorylated, even within the limits of very high phosphorylation rates.

\begin{figure}
\includegraphics[scale=0.7]{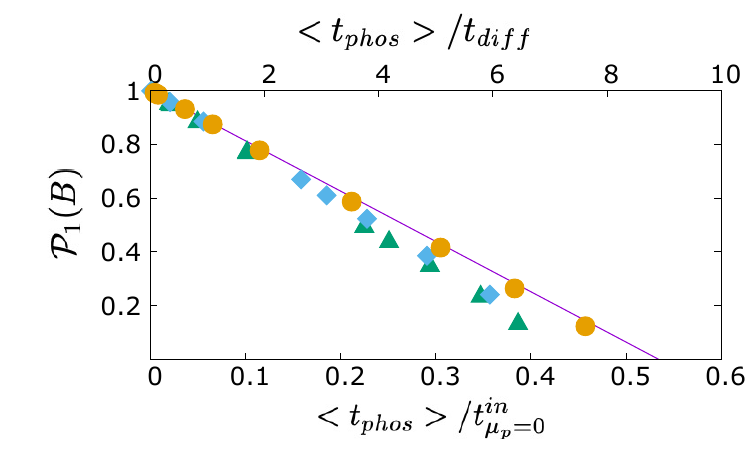}
\caption{\label{fig:fig_6bis_article}
Probability that a protein will be mono-phosphorylated in model $B$ as a function of the average time required for the process to reach completion (success). The top x-axis corresponds to normalization by the diffusion time $t_{diff} = R_a^2/8D$, while the bottom scale is a function of the time actually spent in the nanocluster before desorption, see Eq. (\ref{eq:def_time_in_disk}). Due to the non-zero desorption rate, $\mu_d>0$, not all proteins are phosphorylated. The characteristic length $l_d = \sqrt{D/\mu_d}$ is therefore finite ($D_1= D_2=D$). The solid line corresponds to the analytical formula (\ref{eq:t_mean_as}) and (\ref{eq:P1B_as}) calculated to the first order in $1/\mu_p$ when $\mu_p$ is large, and it is a priori valid only for $<t_{phos}>/t_{diff} \ll 1$ (see Appendix \ref{Proba_versus_mean_time}). The symbols $\bullet$ correspond to the results of a numerical simulation where the activation zone is circular with size $R_a$ and $R_a/l_d =0.56$. For the other symbols, the activation zone has the shape of a small square of 16 elementary cells. The symbols $\blacklozenge$ correspond to a homogeneous system with a constant phosphorylation rate $\mu_p$, while the symbols $\blacktriangle$ correspond to the case where cells dynamically alternate between an active state and an inactive state (dynamic disorder). }
\end{figure}

As mentioned earlier, an ideal single-molecule experiment measures the probability of mono-phosphorylation, ${\mathcal{P}}_1(B)$, as well as the average time $<t_{mean}>$ its take for a protein to be phosphorylated. Each of these two quantities depends on the average phosphorylation rate $<\mu_p>$, which is one of the parameters of the problem. Consequently, the parametric curve
\begin{equation} \label{eq:parametric-plot}
\left(<t_{mean}>, {\mathcal{P}}_1(B)\right)
\end{equation}
makes it possible to avoid determining this parameter in making predictions. As seen in Fig. \ref{fig:fig_6bis_article}, this graph gives an almost straight line we now interpret.

As a reference, Appendix \ref{Proba_versus_mean_time} provides the leading behavior of both quantities in the high phosphorylation regime. Each term varies as $1/\mu_p$, so that the graph shows a straight line consistent with the stochastic numerical data for sufficiently large radii $R_a$, even in the low phosphorylation regime. This graph shows, however, small deviations for small, dynamically disordered activation disks with a square shape (see insets of Fig. \ref{fig:fig6}) only in the low phosphorylation regime. As for the problem in \cite{krapivsky2017}, one expects, therefore, that the asymptotic forms are independent of the circular geometry considered here. As a consequence of this parametric plot, changing the effective phosphorylation rate, for example, by modifying the residence time of kinases, allows the graph to be traversed, assuming that the desorption rate $\mu_d$ of the protein remains unchanged.

A second consequence of the approach proposed here is to show how phosphorylation clusters enable or prevent crosstalk between different signaling pathways by renormalizing effective rate constants. In parallel with single-molecule kinetics \cite{Ninio:1987mq,Kou:2005en}, the inverse of the first moment of the waiting time distribution for a single protein gives an effective cluster-size-dependent phosphorylation rate for an ensemble average measurement. In the large phosphorylation regime with $\mu_p \gg \mu_d$, this rate, $\mu_{p,eff}$, is proportional to $\mu_p$ renormalized by a function of $R_a/l_d$ with
\begin{equation}
\mu_{p,eff} \approx 2 \mu_p \frac{K_0^2(R_a/l_d)}{Q(R_a/l_d)}
\end{equation}
as given in Appendix \ref{Proba_versus_mean_time}. This rescaling takes the value 1 when $R/l_d$ is large but approaches 0 in the other limit and has a simple physical interpretation. When the desorption rate is large, $R/l_d \gg 1$, phosphorylation takes place on a single site without diffusion. The rate is therefore $\mu_p$. The other limit, equivalent to that of a large diffusion coefficient, implies that the protein, although returning many times to the activation disk before desorbing, spends very little time in this disk and, as a result, its phosphorylation rate decreases.

\subsubsection{Crossover between model A and B}
So far, we have considered the special case of a homogeneous diffusion coefficient. In the context of cell adhesion, this assumption is not necessarily always valid.

The above results can be generalized to the following situation: For $0 \le r \le R_a$, let $D_1$ be the diffusion coefficient, while it is $D_2$ outside, with $D_2>D_1$. Clearly, excursions away from the activation disk are favored, and the mortal protein is expected to desorb more frequently before returning in its round trips. Using the results of Appendix \ref{sectionA1-2}, Fig. \ref{fig:Diff_Noteq} clearly shows the effect of increasing $D_2$ relative to $D_1$, where the probability of phosphorylation crosses over between the two limiting cases, model A and model B.

\begin{figure}
\includegraphics[scale=0.7]{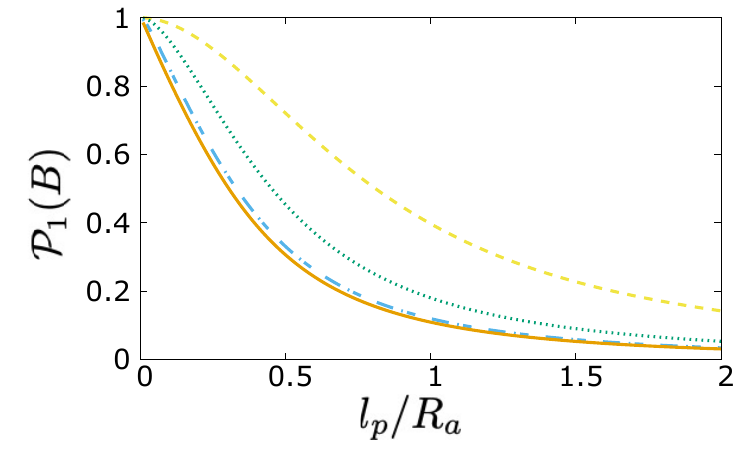}
\caption{\label{fig:Diff_Noteq} Crossover between model A and B. Probability of a protein being phosphorylated once as a function of the parameter $l_p/R_a$, with $l_p = \sqrt{D_1/\mu_p}$. For reference, the upper dashed curve corresponds to $D_2=D_1$ (model B) and the continuous bottom one is for absorbing boundary conditions (model A) at the boundary of the activation disk. The two other dashed curves correspond to $D_2 = 10D_1$ and $D_2 = 100 D_1$ (bottom) and illustrate how the system crosses over from model A to B.}
\end{figure}

\section{The multi-phosphorylation case}

\begin{figure*}
\includegraphics[scale=0.5]{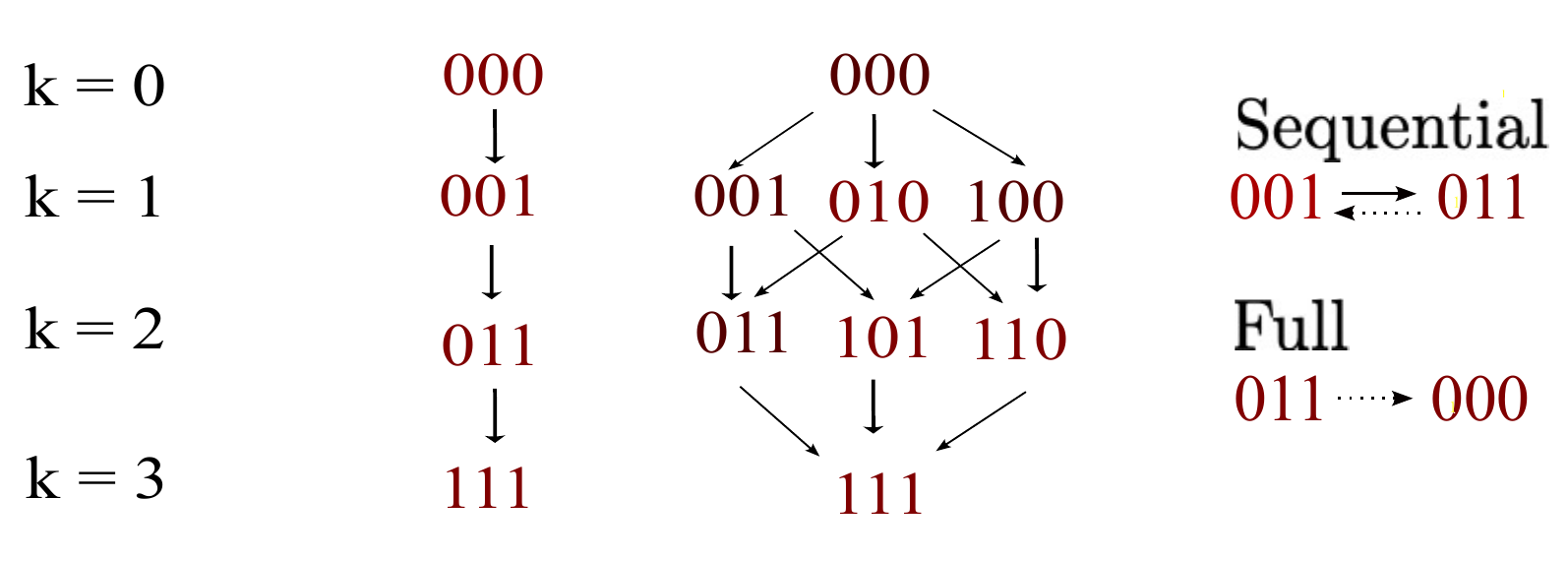}
\caption{\label{fig:fig8}Phosphorylation networks. A protein can be phosphorylated on any or all of its $n$ residues ($n=3$ here) as indicated by the arrow. The left-hand column shows the total number $k$ of phosphorylated residues at each level of the hierarchy. The network in the middle is a sequential network, while the network on the right is a random network. For the random network, there are $C_n^k$ possibilities at each level for a total of $2^n$ possible states and $2^{n-1}n$ phosphorylation reactions. Dephosphorylation can be added either sequentially by modifying each edge (arrow) to make it bidirectional, e.g., $(001) \leftrightarrow (011)$, or by adding a complete dephosphorylation for each residue, for example, $(011) \rightarrow (000)$, both models with a $\mu_{dp}$ rate.}
\end{figure*}

We consider a protein that can be phosphorylated on an arbitrary number of residues and take $D_2=D_1=D$. Dephosphorylation must be taken into account and will be addressed in a second subsection. We introduce a threshold $n$ as the minimum number of phosphorylated residues to serve as a criterion for process completion. If a protein is deposited at time $t=0$ in the disk, the process is completed in two cases:
\begin{enumerate}
\item Model A (immortal protein): The protein exits the disk without having been phosphorylated on $n$ residues (failure), or the protein is phosphorylated on $n$ residues without having crossed the boundary (success).
\item Model B (mortal protein): The protein is phosphorylated on $n$ residues (success) or desorbs from the membrane (failure).
\end{enumerate}
\subsection{Simple multi-phosphorylation networks}

Whatever the model (A or B), multi-phosphorylation can be either sequential or random. In the former case, the sequence of events follows an order where, for example, residue 1 must be phosphorylated for residue 2 to be phosphorylated, and so on. In an alternative version, the order in which the residues are phosphorylated is irrelevant, and the networks are said to be random, as shown in Fig. \ref{fig:fig8}.

\begin{figure}
\includegraphics[scale=0.7]{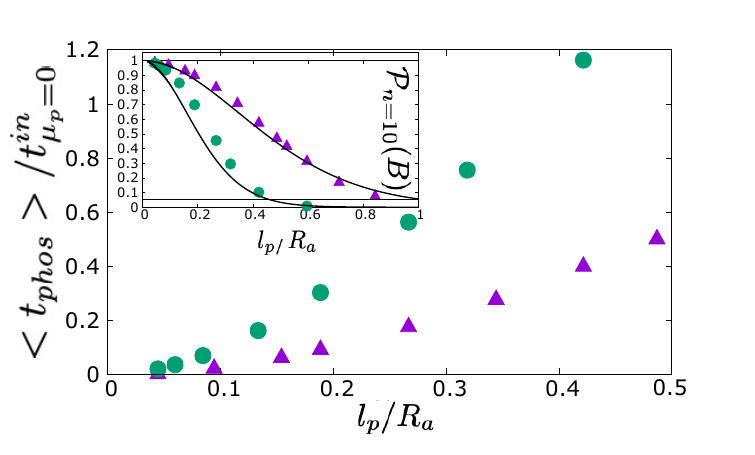}
\caption{\label{fig:fig9} Mean time for a protein to be phosphorylated on $n=10$ residues (normalized by $t_{\mu_p =0}^{in}$ as defined in (\ref{eq:def_time_in_disk})). The upper curve (circles) corresponds to the sequential network and the bottom one to the random network (up-triangles). Inset: Probability for a protein to be phosphorylated on $n=10$ residues for a sequential (circle, bottom curve) and random (up-triangle, upper curve) network. Numerical data fit the two analytical curves calculated in the text with error due to finite-size effects (see (\ref{eq:tous_phospho})). Data correspond to mortal proteins with $\mu_d=5 \times 10^{-4}$ or $l_d/R_a = 1.89$. }
\end{figure}
Each network behaves differently, since the phosphorylation propensity is constant in the sequential case, whereas it varies at each level of the hierarchy for the random network. For example, in the random case, to go from level $k=0$ (no phospho-residues) to level $k=1$ (1 phosphorylated residue among $n$), this propensity is $n \mu_p$. This difference translates into a higher probability for the random network with a shorter time, given that the process is completed as soon as $n$ phospho-residues are reached (success). This is illustrated in Fig. \ref{fig:fig9} for a mortal protein that can exit and re-enter the activation disk before desorption (failure). Thus, a sequential network has a lower probability of being fully phosphorylated, and this takes a longer time than for a random process.

To calculate this probability, let us first define the propensity function \cite{Erban:2009aa} to go from level $n-k$ to level $n-k+1$ of the hierarchy. The probability of a site being phosphorylated per unit time is $\mu_p$. So the probability of one of the residues being phosphorylated per unit time is the sum of these probabilities, i.e., $(n-k)\mu_p$, and the total propensity is
\begin{equation}
P(n-k \rightarrow n-k+1) ={\mathcal{P}}_1(k \mu_p) \neq k {\mathcal{P}}_1(\mu_p)
\end{equation}
Consequently, multi-phosphorylation is not additive, and because ${\mathcal{P}}_1$ depends on the diffusion constant $D$, diffusion induces some cooperative effects between the residues. For simplicity, consider model A. Let ${\mathcal{P}}_1(\mu_p)$ be the probability (before escape) if it has only one possible phosphorylation site. Then, as all residues play the same role, the protein has a probability $n \times (1/n) {\mathcal{P}}_1(n \mu_p)$ of being phosphorylated at at least 1 site before exiting.
The probability to exit the disk with exactly one site among the $n$ is therefore
\begin{equation}
{\mathcal{P}}_1(n \mu_p) (1 - {\mathcal{P}}_1((n-1) \mu_p))
\end{equation}
with a probability $1 - {\mathcal{P}}_1(n \mu_p)$ of escaping without being phosphorylated. The process is the same at each step of the hierarchy $k=1, 2, \ldots, n$ phosphorylated residues. The probability of escaping with exactly two phosphorylated residues is
\begin{equation} \label{eq:multi_phospho_reasoning_2}
{\mathcal{P}}_1(n \mu_p) {\mathcal{P}}_1((n-1) \mu_p) (1 - {\mathcal{P}}_1((n-2) \mu_p))
\end{equation}
and for exactly all $n$ phosphorylated residues
\begin{equation} \label{eq:tous_phospho}
{\mathcal{P}}_n ={\mathcal{P}}_1(n \mu_p) {\mathcal{P}}_1((n-1) \mu_p) \ldots {\mathcal{P}}_1((n - n + 1) \mu_p)
\end{equation}
We verify
\begin{equation} \begin{aligned}
1 &- {\mathcal{P}}_1(n \mu_p) + {\mathcal{P}}_1(n \mu_p) (1 - {\mathcal{P}}_1((n-1) \mu_p)) \\
&+ {\mathcal{P}}_1(n \mu_p) {\mathcal{P}}_1((n-1) \mu_p) (1 - {\mathcal{P}}_1((n-2) \mu_p)) \\
& + \ldots + {\mathcal{P}}_1(n \mu_p) {\mathcal{P}}_1((n-1) \mu_p) \ldots {\mathcal{P}}_1(1 \times \mu_p)\\& = 1
\end{aligned}
\end{equation}

\begin{figure}
\includegraphics[scale=0.7]{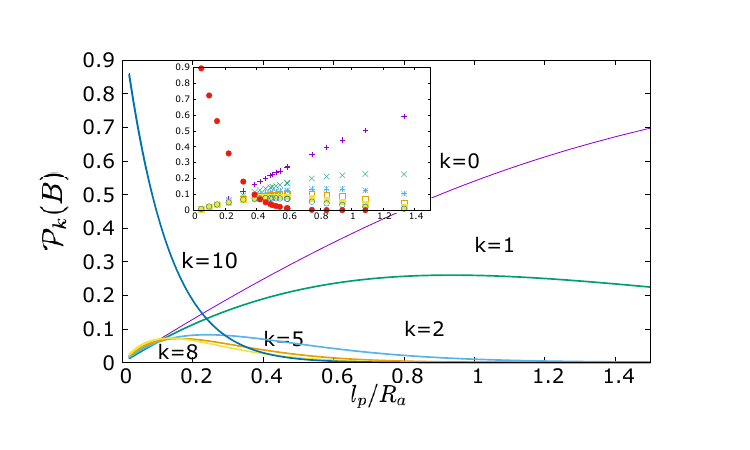}
\caption{\label{fig:proba_n}
Probability of exiting the disk for the first time with $k=1,\ldots,5$ or $10$ phospho-residues ($R=3, \mu_d=0.0$, immortal proteins). The phosphorylation network is random as in Fig. \ref{fig:fig6} ($n=10$). For high phosphorylation rates, $l_p/R \ll 1$, all residues are phosphorylated before crossing the boundary (see $k=10$). Inset: same probabilities from stochastic simulations ($k=10$: red (dark gray) circles, $k=0$: $+$, $k=1$: $\times$, etc.). }
\end{figure}

Using previous results for ${\mathcal{P}}_1$, we can calculate the probability that the protein is phosphorylated on $k=0, 1, 2, \ldots, n$ residues. This is shown in Fig. \ref{fig:proba_n} for model A and in Fig. \ref{fig:fig9} (inset) for model B. In both cases, the system behaves as an all-or-nothing switch as the scaling parameter $l_p/R_a$ is varied. At a sufficiently high rate of phosphorylation, either all of the $n=10$ residues are phosphorylated, or none of them are. The size $R_a$ of the nanoclusters cannot select the number of phosphorylated residues. Because the propensity for phosphorylation varies in the hierarchy, the time necessary for the process to go from level $k=0$ to full phosphorylation with $k=n$ is much faster in the random multi-phosphorylation case than in the sequential model, as shown in Fig. \ref{fig:fig9}.

\begin{figure}
\includegraphics[scale=0.7]{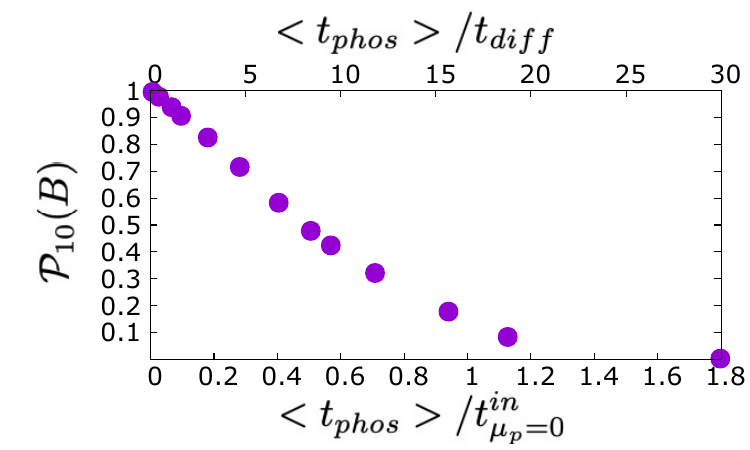}
\caption{\label{fig:fig10_bis} Probability that a protein is phosphorylated on $n=10$ residues as a function of the mean time necessary to reach completion (success).
As in Fig. \ref{fig:fig_6bis_article} and for comparison, the time scale is normalized in two ways. The top x-axis corresponds to normalization by the diffusion time $t_{diff} = R_a^2/8D$, while the bottom scale is a function of the time actually spent in the nanocluster before desorption, see Eq. (\ref{eq:def_time_in_disk}). Data correspond to $R_a/l_d = 1.78$ with $R_a=3$.}
\end{figure}

As before and to compare with Fig. \ref{fig:fig_6bis_article}, Fig. \ref{fig:fig10_bis} shows the graph of the probability that a protein is phosphorylated on $n=10$ residues as a function of the average time to complete the process. Clearly, the linear nature of the behavior is still present within the limits of the phosphorylation rates. However, as expected, the characteristic times for the kinetics are significantly slower in the case of multi-phosphorylation.

\subsection{Multi-phosphorylation with phosphatases}

\begin{figure}
\includegraphics[scale=0.7]{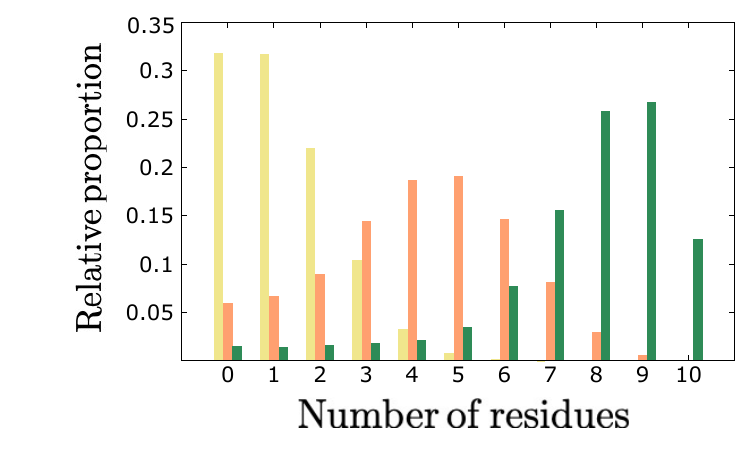}
\caption{\label{fig:proba_surprise} Normalized histograms of the number $n$ of phosphorylated residues when the protein first crosses the disk boundary ($n=10$). The sequential dephosphorylation rate is set at $\mu_{dp} = 0.05$ with three phosphorylation rates ($\mu_{p}=0.25$ (green - dark gray), $\mu_{p}=0.05$ (orange - medium gray) and $\mu_{p}=0.01$ (yellow - light gray)), corresponding to $l_p/R_a = 0.084, \,0.18,\, 0.42$. In this model, the dephosphorylation rate selects the number of phosphorylated residues, since the histograms are either centered on small $n$'s ($\mu_{p} \ll \mu_{dp}$) or on large $n$'s ($\mu_{p} \gg \mu_{dp}$)).}
\end{figure}

Dephosphorylation, a function orchestrated by phosphatases, can be easily included in the model. The first way to do this is to assume that phosphorylation sites in the disk $R_a$ are also dephosphorylation sites, so that a protein visiting the $R_a$ disk by diffusion is subject to a chemical reaction of the type
\begin{equation}\label{eq:seq}
(\ldots, 0_i, \ldots) \xrightarrow[]{\mu_p} (\ldots, 1_i, \ldots) \xrightarrow{\mu_{dp}} (\ldots, 0_i, \ldots)
\end{equation}
with rates $\mu_p$ and $\mu_{dp}$. Here, this scheme is termed   sequential, because each phospho-reaction has its own inverse. To parallel the McKeithan proofreading mechanism, another scheme, referred to here as   total, includes complete dephosphorylation at all levels of the hierarchy:
\begin{equation}\label{eq:full}
(\ldots, 1_i, \ldots 1_j, \ldots) \xrightarrow{\mu_{tdp}} (0_1,\ldots, 0_i, \ldots,0_j, \ldots, 0_n)
\end{equation}
It is instructive to start with the first scheme by generalizing model A. The process is completed if the number of phosphorylated residues reaches a maximum threshold, or if the protein crosses the boundary. In the latter case, the number of phospho-residues is retained. Using (\ref{eq:seq}), we can plot the histograms shown in Fig. \ref{fig:proba_surprise}, where all results correspond to the same phosphorylation rate $\mu_p$. For this value of $\mu_p$, more than 90\% of proteins are fully phosphorylated if dephosphorylation is not included in the model. Clearly, dephosphorylation, the propensity of which increases with the number of phosphorylated residues, enables histograms to be shifted leftward, and therefore the disk determines a specific number of residues to be phosphorylated in the hierarchy $k=0, \ldots, n$. In this case, the size $R_a$ is a selection factor, but as seen below, this case is marginal.

\begin{figure}
\includegraphics[scale=0.7]{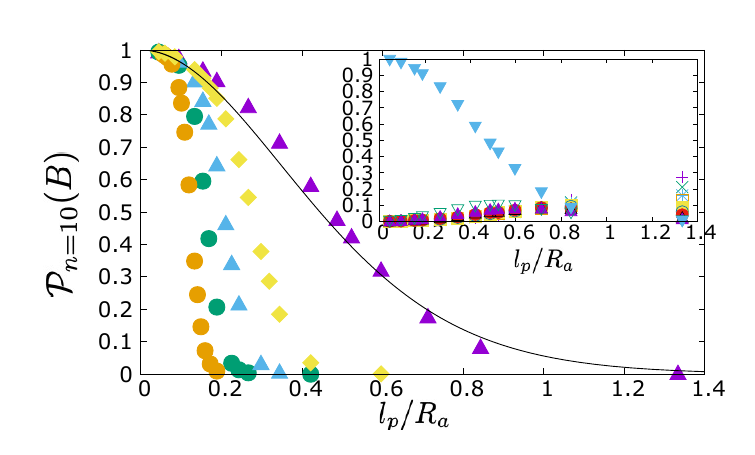}
\caption{ \label{fig:fig12} Probability for $n=10$ residues to be phosphorylated for different sequential dephosphorylation rates ($\mu_{dp} = 0.00, \,0.01, \,0.025, \, 0.05 ,\, 0.1$ and $\mu_d = 5 \times 10^{-4}$). The line corresponds to no dephosphorylation rate $\mu_{dp} = 0$ and serves as a reference. The higher the rate of dephosphorylation, the more the curve is sigmoid-shaped. Inset: Probability of the residue number $k$ being phosphorylated ($\mu_{dp}=0$). In the high phosphorylation limit, $l_p/R_a=1$, the protein is fully phosphorylated on $k=10$ residues (blue down triangles). In the other limit, the protein is moderately phosphorylated at all of the residues ($k=0$, symbol $+$). }
\end{figure}

\begin{figure}
\includegraphics[scale=0.7]{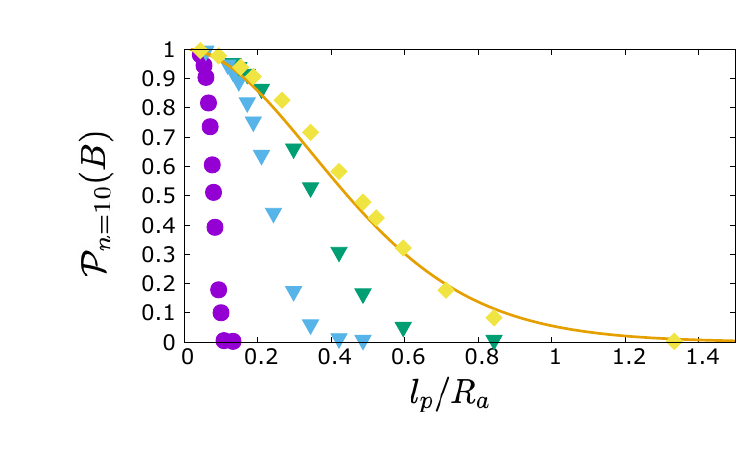}
\caption{\label{fig:fig13} Probability for $n=10$ residues to be phosphorylated for different total dephosphorylation rates ($\mu_{tdp} = \,0.0, \,0.01, \,0.05, \,0.1$ (magenta circles - dark gray)), see Eq. (\ref{eq:full}). The line corresponds to no dephosphorylation rate $\mu_{tdp} = 0$ with yellow (light gray) lozenges and serves as a reference. As in Fig. \ref{fig:fig12} for sequential dephosphorylation, the higher the rate of dephosphorylation, the greater the variations in the curve, so that the system switches abruptly from full phosphorylation to zero by varying the parameter $l_p/R_a$. }
\end{figure}

\begin{figure}
\includegraphics[scale=0.7]{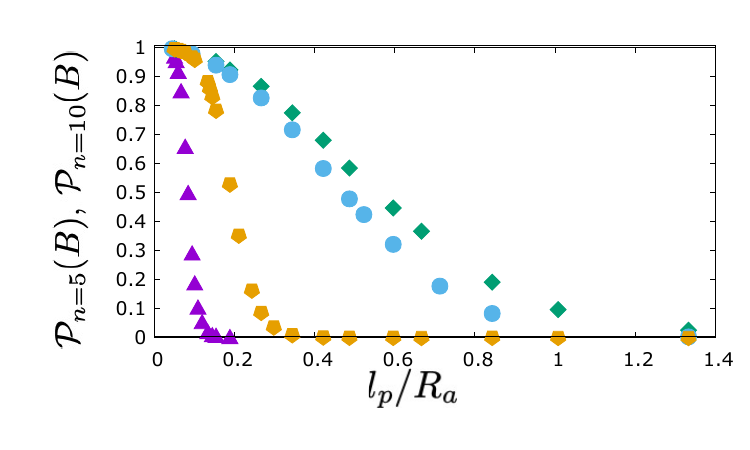}
\caption{\label{fig14_article} Probability for $n=5$ residues all to be phosphorylated (upper green losange symbols, the maximum number of residues is $n=5$). For reference, this probability, but for a maximum number of $n=10$ residues also all phosphorylated, is shown as blue circle symbols (same as in the inset of Fig. \ref{fig:fig9}). The orange pentagon and magenta triangle symbols correspond to a dephosphorylation rate $\mu_{dp} = 0.1, \,1$ for a maximum number of $n=5$ also all phosphorylated.}
\end{figure}

Taking into account the loop effects of model B changes this picture. For the same dephosphorylation scheme (\ref{eq:seq}), Fig. \ref{fig:fig12} shows the effect of sequential dephosphorylation for a protein to be completely phosphorylated. In this case, dephosphorylation mainly affects the subset that goes back and forth, since for $l_p/R_a \ll 1$, proteins are phosphorylated very rapidly and the tangent to the curve remains horizontal. Increasing $\mu_{dp}$ very quickly, the curves take on the shape of an inverted sigmoid, with a threshold effect for a value of parameter $l_p/R_a$ below which proteins are completely phosphorylated. The effect is the same for the total phosphorylation scheme of Eq. (\ref{eq:full}), see Fig. \ref{fig:fig13}, and is almost independent of the maximum number of residues, since decreasing the maximum number of residues from $n=10$ to $n=5$ produces the same effect, as seen in Fig. \ref{fig14_article}. We therefore conclude that this is a robust property for networks.

\begin{figure}
\includegraphics[scale=0.7]{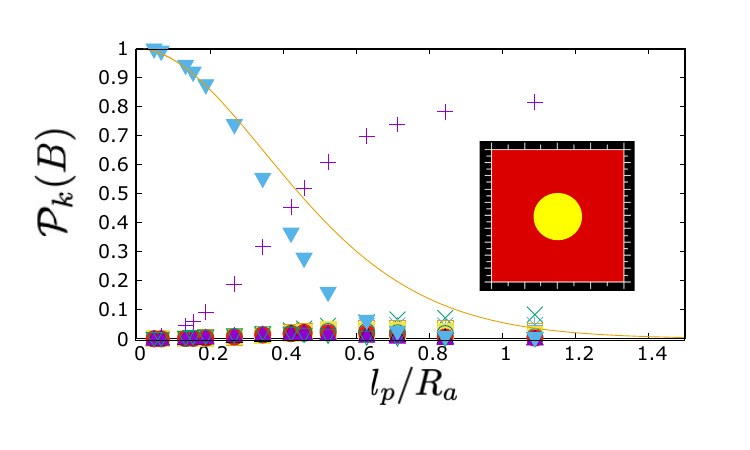}
\caption{\label{fig:fig15}Probability of phosphorylating residue $k$ for a protein with a maximum number of $n=10$ residues ($k=10$ corresponds to $\blacktriangledown$ and $k=0$ to $+$ symbols). For this simulation, the protein can only be (multi)phosphorylated in the activation disk of radius $R_a$, whereas it can only be dephosphorylated (total dephosphorylation) outside it (area colored red). The rate of $\mu_{tdp}$ dephosphorylation is 5 times greater than the maximum $\mu_{dp}$ in Fig. \ref{fig:fig12} and only the tail of the distribution without dephosphorylation (solid curve) is affected by dephosphorylation.
}
\end{figure}

To highlight the importance of the local nature of post-translational modifications in obtaining switch-like responses, consider Fig. \ref{fig:fig15}. In this case, we depart from the previous rule and introduce phosphatases only outside the activation circle, so that only proteins localized outside the disk can be dephosphorylated in a single reaction (total dephosphorylation). Even for very high rates of dephosphorylation, the response is gradual, not abrupt, because dephosphorylation only affects the tail end of the distribution where proteins are most subject to round-tripping.

\section{Potential biological relevance and conclusion}

The ability to mono- or multi-phosphorylate proteins is one of the essential features of cell signaling networks, and this work proposes a reaction-diffusion approach to study this problem in the context of adhesion. To calculate these probabilities, we have introduced a dimensionless parameter that relates individual properties of each protein, such as the diffusion coefficient, to the spatiotemporal average enzymatic activity in an activation zone, as well as to its characteristic size. This probability is highly dependent on this parameter and, depending on phosphatase activity, can even lead to abrupt or almost switch-like biochemical behavior, in contrast to more gradual variations. Introducing this effective parameter together with characteristic length and time scales is, in our view, a way of rationalizing the processes that control molecular self-assembly. It allows also to define effective nanoscopic rates to rationalize crosstalks between different signaling pathways. 

It is likely that multi-phosphorylation plays a different role than mono-phosphorylation in the stability of molecular assemblies and biomolecular condensates that the cell regulates, among others, in the immediate vicinity of the adhesive zones. In analogy to what was shown for the SH3-Proline-rich region \cite{Li:2012ud}, phase separation following multi-phosphorylation could be driven by the assembly of multivalent proteins. Multi-phosphorylation of residues, larger than 3-4, will support a tunable multivalency on a protein of interest that is essential for phase stability. Moreover, this phosphorylation-dependent multivalency could also provide greater local molecular rigidity and thus be implicated in a new type of mechano-sensing. In terms of mechanobiology, control of multi-phosphorylation is essential to activate a specific signaling pathway through p130Cas \cite{Sawada:2006lr}.

This work demonstrates that simple principle models can account for a fine-tuned regulation of mono- versus multi-phosphorylation, taking into account the effect of the environment and involving biochemical switches working with a small number of molecules. From a biochemical point of view, the main hypothesis is that phosphorylation is a locally controlled phenomenon, as for Tyrosine residues, whereas it seems to be more diffuse on a cellular scale for Ser-Thr ones because of specificity or abundance reasons \cite{Ubersax_2007}.

Controlling the enzymatic activity within adhesive structures is obviously the nerve center of any biochemical control, and $\mu_p$ is interpreted here as an effective rate constant that may depend on a "control variable," say ${\mathcal{C}}$, such as a change of conformation or a local concentration of co-activators. Triggering a signaling pathway implies, however, some kind of time averaging. As usual \cite{McCammon:1981nx,zwanzig1990,Gopich:2016bq}, if ${\mathcal{C}}(t)$ varies in time and if its fluctuations are sufficiently fast with respect to the diffusion time, we have shown using spatiotemporal variations of $\mu_p({\mathcal{C}})$ among elementary cells that $\mu_p$ can be replaced by the time-averaged effective rate, which reflects the dynamics of enzymatic activity. Changes in this dynamics by adding telegraphic (dichotomous) noise  can have many origins, for example, changes in the fraction of time an enzyme spends in its active conformation or changes in how proteins are recruited. Since most of the kinases are cytosolic proteins recruited from the cytosol, Src-kinase being a paradigmatic example, changing their mode of recruitment by molecular engineering will bias their dynamics with a potential strong effect on their effective enzymatic activity. Thus, our model, which demonstrates abrupt variations in phosphorylation, represents a first step towards a potential explanation of the major changes in molecular assemblies demonstrated in \cite{Kerjouan:2021ml} where opto-Src lacks its major anchor domain \cite{Le-Roux:2016zc,Le-Roux:2019cd}. Note that this approach, which focuses on the dynamics and the labile character of a single enzyme to induce changes in its function, has also been proposed in a parallel context in \cite{Seze:2025ik} for a GTP-ase.

Triggering a signaling pathway also implies some kind of local space averaging, and the second principle on which we have based our analysis is that this sensing specificity occurs through effective parameters to allow crosstalk. One signaling pathway interferes with another, as it reads biophysical parameters influenced by the other. The integration of their relative dynamics parameters should control their level of crosstalk and thus drive the system from model A to B with a switch-like behavior. The third principle highlighted in this work is time. To test proteins with nanoclusters having a short lifetime, multi-phosphorylation must occur rapidly, with characteristic times that depend on the ensemble of phosphorylatable substrates available.

Following on from works on signal transduction \cite{Koivomagi:2013bl,Anselmi:2020xe}, this work suggests that it is possible to induce two different signaling events by playing on the spatial distribution and spacing of phosphorylation sites on a single substrate, leading to structural and temporal changes in the downstream molecular complexes associated with these substrates.

As studied here, however, cooperative effects that translate into parameter-dependent abrupt changes, enabling fine regulation by multi-phosphorylation, must include phosphatase networks, whose effects are less often discussed in {\it {in vivo}} studies than those of kinases.

\begin{acknowledgments}
O.D. is partly supported by the Agence Nationale de la Recherche PRC program (ANR-22-CE13-0046). We thank G. Gogl for useful discussions and an anonymous referee to draw our attention to the problem $D_2 \neq D_1$. We would also like to thank the second referee for his comments, which helped us improve the content of this paper, and for drawing our attention to Ref. \cite{SHOUP1982237}.
\end{acknowledgments}

\appendix

\section{Stochastic simulations} 
\label{Appendix0}

The stochastic counterpart to the Fokker-Planck equation is the Master equation, which we solve numerically using the Gillespie SSA algorithm. We follow the approach of \cite{erban} with an updated list of references from \cite{Cao:2014qy,Cao:2019mw}. The method is also described in \cite{fourcade:hal-02465425} and is a spatial stochastic simulator for chemical reaction networks, see also Ref.  \cite{Andrews_2016}.

The computational domain is divided into a $25 \times 25$ grid of compartments to simulate diffusion as Markov jumps between compartments and reactions within each compartment. Each compartment is treated as spatially homogeneous, but concentrations can differ between them. If $A_i$ is the number of molecules of chemical species A in the $i$-th compartment, the diffusion-reaction process is described by the following system of chemical reactions:

\begin{equation}
\ce{ A_1 <=>[D/h^2][D/h^2] A_2 <=>[D/h^2][D/h^2] A_3 \ldots A_{N-1} <=>[D/h^2][D/h^2] A_N }
\end{equation}

Here, $h$ is the size of one compartment and $D$ is the diffusion constant in the continuum case. 
Typical simulations were run with $D/h^2 = 0.1$, which provides a reference rate for a numerical experiment \cite{erban}.

At each time step, the algorithm computes a propensity matrix to determine the time and type of the next chemical reaction. To save computational time, the propensity matrix for diffusion is updated at each step using a complex double-linked list of structures. The program can process any binary reaction with local reaction constants and generates the in silico statistics used in the text. This approach handles the combination of different states directly without approximation. For instance, the case $n=10$ requires $1025$ different protein phosphorylation classes with $6144$ reactions. Because this method explicitly includes protein diffusion, it naturally stands out from other well-mixed approaches. Trajectory animations can be generated at each time step using the Dislin library \cite{dislin}.

Since the diffusion domain in a numerical simulation is necessarily finite, we introduce an absorbing boundary on a circle with a radius larger than that of the activation disk. A counter is used to track events where the protein makes excursions wider than the numerical diffusion domain allows (event (4) in Fig. \ref{fig:fig2} on the outer circle). These events are discarded when averaging over different realizations, which introduces a controllable numerical error. The other parameters are the same as in the text, with averages taken over $20 \times 10^3$ to $10^6$ samples, depending on the case.

\section{Propagator} 
\label{Appendix1}
Here we demonstrate (\ref{eq:proba_one_model_A}) and (\ref{eq:proba_one_model_B}).
We place a protein at $r=r_0, \, 0 \le r_0 \le R_a$, at $t=0$. Calculating the probability of the protein being phosphorylated is equivalent to calculating  
\begin{equation} \begin{aligned}
&2 \pi \mu_p \int_0^R dr\, r \int_0^\infty dt \, {\mathcal{P}}_1(r,t) \\
&= 
2 \pi \mu_p \int_0^R dr\, r \tilde {\mathcal{P}}_1(r,s=0) 
\end{aligned}
\end{equation} 
where  ${\mathcal{P}}_1(r,t) $ is the survival probability and  $\tilde {\mathcal{P}}_1(r,s=0) $ its Laplace transform. We average the results over $r_0$ and we separate into regions $1, \, 2$ with the definitions ($D_{i=1,2}$ are diffusion coefficients)
\begin{equation}
\begin{aligned}
&\text{Region 1: } D_1, \, 0 \le r \le R_a \\
&\text{Region 2: } D_2, \, R_a < r \le \infty
\end{aligned}
\end{equation}
where the Laplace transform is solution of:
\begin{equation}
\frac 1r \frac{\partial}{\partial r} r \frac{\partial \tilde {\mathcal{P}}_1}{\partial r} - q_{1}(s)^2 \tilde {\mathcal{P}}_1 = -\frac{1}{2 \pi r_0 D_1} \delta (r-r_0) 
\end{equation}
with 
\begin{equation}
 q_{1,2}^2(s) = \frac{\mu_{1,2} +s}{D_{1,2}}
\end{equation} 

In the notations of the text $\mu_1 = \mu_p + \mu_d \approx \mu_p$ (phosphorylation rate, desorption rate) and $\mu_2 = \mu_d$. We have 
\begin{equation}
q_1(s=0) = 1/l_p \quad q_2(s=0) = 1/\lambda
\end{equation}
We  look for a solution in the form 
\begin{equation}
\begin{aligned} 
&A_1 I_0(q_1(s) r) \text{ for } r < r_0\\
&A_2 I_0(q_1(s) r) + B_2 K_0(q_1(s)r) \text{ for } r_0 \le r < R_a\\
&B_3 K_0(q_2(s)r)\text{ for } r \ge R_a
\end{aligned}
\end{equation}
with boundary  conditions: 
\begin{equation}
\eval{\frac{d \tilde {\mathcal{P}}_1}{dr}}_{r =r_0^+} - \eval{\frac{d \tilde {\mathcal{P}}_1}{dr}}_{r =r_0^-} = - \frac{1}{2 \pi D_1r_0}
\end{equation}
where probabilities are continuous in $r=R_a$.

It will be useful to define:
\begin{equation} \begin{aligned}
Z_1 & = \frac{D_2}{D_1}q_2(s) I_0(q_1(s) R_a) K_1(q_2(s)R_a) \\
& + q_1(s) K_0(q_2(s)R_a) I_1(q_1(s)R_a)
\end{aligned}
\end{equation}
\begin{equation}
\begin{aligned} 
Z_2 &= q_1(s) K_0(q_2(s)R_a) K_1(q_1(s)R_a )\\ 
& - q_2(s) \frac{D_2}{D_1}K_0(q_1(s)R_a) K_1(q_2(s)R_a)
\end{aligned}
\end{equation}
Hence: 
\begin{equation} \label{eq:solution}
\begin{aligned} 
A_1 & = \frac{1}{2 \pi D_1} K_0(q_1(s) r_0) + \frac{Z_2}{2 \pi D_1 Z_1} I_0(q_1(s) r_0) \\
A_2 &= \frac{Z_2}{2 \pi D_1 Z_1} I_0(q_1(s) r_0) \\
B_2 &= \frac{1}{2 \pi D_1} I_0(q_1(s) r_0) \\
B_3 &= \frac{1}{2 \pi D_1 R_a Z_1} I_0(q_1(s) r_0)
\end{aligned}
\end{equation}

It's easier to do the mean value on $r_0$ first:
\begin{equation}
I(r) = \frac{2}{ R_a^2} \int_0^{R_a} dr_0\, r_0 \tilde {\mathcal{P}}_1(s=0,r)
\end{equation}
and then do the integral over $r$ times $\mu_p$
\begin{equation}
2 \pi \mu_p\int_0^{R_a} dr \, r I(r)
\end{equation}
to find the Laplace transform at $s=0$ of a probability of a protein being phosphorylated 
\begin{equation} \label{eq:Laplace_resu}
{\mathcal{P}}_1(B) = \frac{\mu_p }{q_1^2 D_1} \left [ 1 - \frac{2 q_2}{q_1 R_a Z_1}  \frac{D_2}{D_1} I_1(q_1 R_a) K_1(q_2 R_a) \right] 
\end{equation}
with 
\begin{equation}
q_{1,2}= q_{1,2}(s=0) = \sqrt{\mu_{1,2}/D_{1,2}} 
\end{equation}

\subsection{Limit cases with equal diffusion constant $D_1=D_2=D$}
\label{sectionA1-1}
Take $D_{1,2} = D$ first . We find the limit of problem A (absorption on the disk boundary $r=R$), when $\mu_2 = \mu_d \gg 1$, i.e. $q_2 \gg1$. In this limit   
\begin{equation}
Z_1 \approx q_2 I_0(q_1R_a) K_1(q_2R_a) \,, \quad q_2 \gg 1
 \end{equation}
 which gives the desired result, since $q_1^2 = \mu_p/D = 1/l_p$.
 \begin{equation} \label{eq:resu_proba_boundary}
1 - \frac{2 l_p}{R_a} \frac{I_1(R_a/l_p)}{I_0(R_a/l_p)} 
 \end{equation}
In this limit, the probability flux across the boundary of the disk gives the probability to exit for the first time at time $t$. Its Laplace transform is 
\begin{equation}
\tilde{\mathcal{F}}_{\mu_p}(s) = \frac{2 l(s)}{R_a} \frac{I_1(R_a/l(s))}{I_0(R_a/l(s))} 
\end{equation}
with 
\begin{equation}
l(s) = \sqrt{D/(\mu_p +s)}
\end{equation}
In the other limiting case where $\mu_d \ll 1$ outside the disk of size $R$, $Z_1 \approx - q_1 I_1(q_1R_a) \ln q_2 \gg 1$ and we find  that the probability is $1$ in the limit where the logarithm is sufficiently large.

Surprisingly, the inverse Laplace transform ${\mathcal{F}}_{\mu_p}(t)$  of $\tilde {\mathcal{F}}_{\mu_p}(s)$ does not seem to be known.
 For $t \rightarrow 0$, i.e. $s \rightarrow \infty$ 
\begin{equation}
{\mathcal{F}}_{\mu_p}(t) \simeq \left( \frac{4D}{\pi R_a^2} \right)^{1/2} \frac{e^{-\mu_p t}}{t^{1/2}}
\end{equation}
We can also calculate the average time $t_{diff}$ it takes for the protein to leave the disk for the first time when $\mu_p = 0$ by calculating $-{\mathcal{F}}'_{\mu_p = 0}(s=0) = R_a^2/(8D)$, which differs by a factor of $2$ from the definition of the non-averaged diffusion time $R_a^2 = 4Dt_{d}$.

\subsection{ Case where $D_2 > D_1$}
\label{sectionA1-2}
We study the case where the diffusion constant $D_2$ outside the disk $0 \le r \le R_a$ is much larger than the one inside the disk $D_1$. In (\ref{eq:Laplace_resu}), take $D_2 \gg 1$, then $q_2 \ll 1$ and $K_1(q_2R_a) \approx 1/q_2R_a$.  With logarithmic corrections $Z_1 = D_2/(D_1R_a) I_0(q_1R_a)$, and $K_1(q_2R_a) \approx 1/q_2R_a$. So we find again the result of Eq. (\ref{eq:resu_proba_boundary}).  The crossover between $D_2=D_1$ (model B) and $D_2 \gg D_1$ (model A) is illustrated in Fig. \ref{fig:Diff_Noteq}. 


\section{Mean time (model A)}
\label{Appendix3}
\renewcommand{\thefigure}{A2-\arabic{figure}}
\setcounter{figure}{0}

This Appendix takes up the previous reasoning on calculating the probability of phosphorylation. We deposit the protein inside the yellow disk at a distance $a <R$ from the center and we calculate the probability that this protein reaches the boundary $r=R$ between $t$ and $t+ dt$. This probability is denoted ${\mathcal{F}}_{\mu_p}(t,r_0)dt$ and its Laplace transform is $\tilde {\mathcal{F}}_{\mu_p}(s,r_0)$. We have 
\begin{equation}
\tilde {\mathcal{F}}_{\mu_p}(s,r_0) = \int_0^\infty e^{-st} {\mathcal{F}}_{\mu_p}(t,r_0) dt
\end{equation}
Consequently, $\tilde {\mathcal{F}}_{\mu_p}(s=0,r_0)$ is the probability of having reached the edge in the interval $[0, \infty[$.   The probability of never having reached the edge at $t$ is then 
\begin{equation}
1 - \int_0^t {\mathcal{F}}_{\mu_p}(t,r_0)
\end{equation}
If the protein has not touched the edge between $[0,t]$, the probability that it will be phosphorylated between $t$ and $t+ dt$ is 
$
\mu_p e^{-\mu_p t} dt
$.
The probability that it will be phosphorylated for the first time between $t$ and $t +dt$ is therefore 
\begin{equation}
\mu_p e^{-\mu_p t} dt ( 1 - \int_0^t du {\mathcal{F}}_{\mu_p=0} (u,r_0) )
\end{equation}
The probability to complete the process at time  $t$ is therefore 
\begin{equation} \label{eq:probability} \begin{aligned}
{\mathcal{P}}(t,r_0) =& \mu_p e^{-\mu_p t} \left [ 1 - \int_0^t du {\mathcal{F}}_{\mu_p=0} (u,r_0) \right] \\
& + {\mathcal{F}}_{\mu_p}(t,r_0)
\end{aligned}
\end{equation}
where we take $\mu_p =0$ in the integrant, because if the protein has touched the edge, it cannot have been phosphorylated.
And after averaging over $0\le r_0 \le R$ 
\begin{equation} \label{eq:probability2}
{\mathcal{P}}(t) = \mu_p e^{-\mu_p t} \left [ 1 - \int_0^t du\, {\mathcal{F}}_{\mu_p=0} (u) \right] + {\mathcal{F}}_{\mu_p}(t)
\end{equation}
To check that this probability is suitably normalized, we note that we have 
\begin{equation}
\tilde {\mathcal{F}}_{\mu_p=0}(s=\mu_p) = \tilde {\mathcal{F}}_{\mu_p}(s=0) 
\end{equation}
because the combination $\mu_p +s$ comes into play. We have, indeed
\begin{equation} \label{eq:propagator}
\tilde {\mathcal{F}}_{\mu_p}(s) = g(x) = \frac 2 x \frac{I_1(x)}{I_0(x)}, \quad x = R_a \left[ \frac{\mu_p + s}{D} \right]^{1/2} 
\end{equation}
In (\ref{eq:probability}), integrating over $t$, we have
\begin{equation}
 \int_0^\infty P(t) dt = \mu_p \int_0^\infty e^{-\mu_p t} dt = 1
 \end{equation}
 because 
 \begin{equation}
 \int_0^\infty \mu_p e^{-\mu_p t} \int_0^t du {\mathcal{F}}_{\mu_p=0} (u ) dt = - \tilde {\mathcal{F}}_{\mu_p}(s=0)
 \end{equation}
The probability of the protein being phosphorylated between $t$ and $t + dt$ is therefore 
\begin{equation} \label{eq:proba_t} \begin{aligned}
& {\mathcal{P}}_{p}(t,l/R) dt = \\ & \frac{1}{\mathcal{Z}} \mu_p e^{-\mu_p t} \left [ 1 - \int_0^t du {\mathcal{F}}_{\mu_p=0} (u) \right] dt
\end{aligned}
 \end{equation}
where ${\mathcal{Z}}$ is a normalization factor. 
\begin{equation}
{\mathcal{Z}} = 1 - g(1/x) \quad x = l_p/R_a
\end{equation}
Its first moment is calculated as 
\begin{equation} \label{eq:first_moment} \begin{aligned}
& <t_{phos}> = \\ &\frac{1}{\mathcal{Z}} \int_0^\infty t \mu_p e^{-\mu_p t}\left [ 1 - \int_0^t du {\mathcal{F}}_{\mu_p=0} (u) \right] dt
\end{aligned}
\end{equation}
 This gives for the first moment 
 \begin{equation} \label{eq:first_moment_resu}
 \begin{aligned}
 &<t_{phos}> = \\
& \frac{1}{\mu_p} \left[1 - g(1/x) + \frac{R}{2l_p} g'(1/x) \right]/(1 - g(1/x)) 
 \end{aligned}
 \end{equation}
If we define the diffusion time as 
 \begin{equation}
 R_a^2 = 8 D t_{diff} \text{ or } \mu_p t_{diff} = R_a^2/ 8 l_p^2
 \end{equation}
 then (\ref{eq:first_moment_resu}) gives the solid curve of Fig. \ref{fig:mean_time}. As the size of the adhesion disk is finite, the time distribution is cut off and, for proteins that are phosphorylated, the mean time is less than $1/\mu_p$.  
 
 Finally, Taylor expanding $g(x)$ for $x\ll 1$ gives  $<t_{phos}> = 1/\mu_p$ as expected.
 
\section{Time spend in the activation disk for a mortal protein}
\label{Appendix:time_spend}
We define a characteristic time $t_{\mu_p =0}^{in}$ as the mean time that a mortal protein spends in the disk before desorbing from the membrane. From the survival probability calculated above, this reference time can be calculated as follows
\begin{equation}
2 \pi \int_0^Rdr \, r \int_0^\infty dt \, {\mathcal{P}}_1(r,t)
\end{equation}
where the Laplace transform  of ${\mathcal{P}}_1(r,t)$ is given in (\ref{eq:solution}).
For  $\mu_p = 0$, one has 
\begin{equation}
Z_1 = 1/(qR_a^2) \quad q = \sqrt{\mu_d/D}
\end{equation}
Hence 
\begin{equation} \label{eq:def_time_in_disk}
t_{\mu_p =0}^{in} = \frac 1 \mu_d \left[ 1 - 2 R_a/l_d I_1(R_a/l_d) K_1(R_a) \right]
\end{equation}
which allows us to plot the renormalized averaged times for a protein to be phosphorylated  on $10$ residues in both cases (sequential and random network) as shown in Fig. \ref{fig:fig9}.

\section{Mean time (model B)}
\label{Proba_versus_mean_time}

 In this appendix, we give the formulas for calculating the average monophosphorylation time in the context of model $B$. Using (\ref{eq:Laplace_resu}) for the Laplace transform, we have 
 \begin{equation}
 <t_{mean}> = - \mu_p   \left[ \frac{1} {\tilde {\mathcal{P}}_1(B)} \frac{\partial {\tilde {\mathcal{P}}_1(B)}}{ \partial s} \right]_{s=0}
 \end{equation}
which gives to leading order in the high phosphorylation regime 
\begin{equation} \label{eq:t_mean_as}
<t_{mean}> \approx   \frac{1}{2 \mu_p} \frac{Q(R/l_d)}{K_0(R/l_d)^2} 
\end{equation}
with corrections of order $1/\mu_p^{3/2}$ and where 
\begin{equation}
Q(x) = K_0\left(x\right){}^2+ K_0\left(x\right) \left( \frac 2 x  K_1\left( x\right)- 
   K_2\left(x\right)\right)+2 K_1\left(x\right)^2
   \end{equation}
As expected,  $<t_{mean}> \approx 1/\mu_p$ in the small desorption limit, but increases significantly in the other limit.   

 In the same limit of large phosphorylation rate, one finds for the probability   
   \begin{equation}
   \label{eq:P1B_as}
{\mathcal{P}}_1(B) \approx 1 - \frac{1}{\mu_p} \left[ 2 \frac{D}{ R_a l_d }\frac{K_1(R_a/l_d)}{K_0( R_a/l_d)} +\mu_d \right]
\end{equation}
so that Eqs. (\ref{eq:t_mean_as}) and (\ref{eq:P1B_as}) give the parametric plot of Fig. \ref{fig:fig_6bis_article}.

\bibliographystyle{apsrev4-2}
%

\end{document}